\documentclass[aps,prd,showpacs,showkeys,twocolumn]{revtex4}
\def\be{\begin{equation}}
\def\ee{\end{equation}}
\def\bea{\begin{eqnarray}}
\def\eea{\end{eqnarray}}
\def\bey{\begin{eqnarray*}}
\def\eey{\end{eqnarray*}}
\def\bml{\begin{mathletters}}
\def\eml{\end{mathletters}}
\def\ba{\begin{array}}
\def\ea{\end{array}}

\def\nn{\nonumber}
\def\to{\rightarrow}
\def\<{\left\langle}
\def\>{\right\rangle}
\def\({\left(}
\def\){\right)}
\def\e{{\rm e}}
\def\cG{{\mathcal{G}}}
\def\cK{{\mathcal{K}}}
\def\cP{{\mathcal{P}}}
\def\C{\hat{C}}
\def\J{\hat{J}}
\def\M{\hat{M}}
\def\B{\hat{B}}
\def\SO{{\rm SO}}
\def\U{{\rm U}}
\def\SU{{\rm SU}}
\def\usp{{\rm usp}}
\def\USp{{\rm USp}}
\def\tr{\,{\rm tr}\,}

\def\ud{u^{\dagger}}

\def\gd{g^{\dagger}}

\def\Psit{\Psi^{T}}
\def\psib{\bar{\psi}}
\def\psid{\psi^{\dagger}}
\def\psis{\psi^{*}}

\def\chib{\bar{\chi}}
\def\chid{\chi^{\dagger}}
\def\chis{\chi^{*}}

\def\D{D\!\!\!\!\slash\;}
\def\Q{\tilde{Q}}
\newcommand{\bx}{\ba{|c|}\hline\ \ \\ \hline\ea}
\newcommand{\bxs}{\ba{|c|}\hline\,\ast\, \\ \hline\ea}

\begin{document}
\preprint{}
\title{Two-color QCD in 3D at finite baryon density}
\author{Gerald V. Dunne}
\affiliation{Department of Physics, University of Connecticut,
Storrs, CT 06269-3046, USA}
\author{Shinsuke M. Nishigaki}
\affiliation{Department of Physics, University of Connecticut,
Storrs, CT 06269-3046, USA}
\date{October 15, 2002}
\begin{abstract}
We study the low energy phase structure of SU(2) gauge theories
in three-dimensional spacetime, at finite baryon density.
The pseudoreality of representations of SU(2) permits an analytic study
of a real
baryon chemical potential, and the restriction to 3D results in a different
global symmetry breaking pattern from the corresponding 4D model
studied previously by Kogut et al.
We find a second-order phase transition separating the
normal phase and the baryon superconducting phase.
The chemical potential dependence of condensates
and baryon density are computed.
We find that the phase structure and the excitation spectrum
are essentially the same as in 4D,
despite the different symmetry groups,
indicating a  universality that is rooted in the properties of
Riemannian symmetric spaces.
\end{abstract}
\maketitle

\section{introduction}
The spontaneous breaking of global symmetries is an essential
feature of realistic gauge theories \cite{miransky}.
Much has been learned about this phenomenon by studying models in
various dimensions.
For example, the
2D Schwinger model and 4D QCD
exhibit the breakdown of discrete or continuous
chiral symmetry, respectively.
For even dimensional spacetime there is a comprehensive theorem that predicts
the surviving part of of the flavor symmetry \cite{vafa}.
For theories in odd dimensional spacetime the situation is rather different, primarily because there is no chirality and the parity transformation acts differently \cite{roman}. In particular, models in 3D spacetime, while different from 4D theories, provide excellent testing grounds for our understanding of dynamical symmetry breaking. Yet even here the picture is still not completely clear.
The generic symmetry breaking pattern in 3D has been described for Abelian theories in \cite{poly}.  In large $N_F$ QED$_3$ there are indications of spontaneous breaking of the global flavor symmetry \cite{pisarski1,stam}. Studies of the Schwinger-Dyson and gap equations for QED$_3$ in the large $N_F$ limit predict that spontaneous symmetry breaking only occurs for $N_F$ below a critical number of flavors $N_F^{\rm crit}=32/\pi^2\approx 3$ \cite{appelquist,cristina}.
These predictions have numerical support from lattice simulations of 3D QED \cite{dagotto}.
On the other hand, other 3D QED Schwinger-Dyson analyses \cite{pennington} predict symmetry breaking for all $N_F\geq 2$, a result which is consistent with the renormalization group arguments in \cite{pisarski2}.

Another approach to the symmetry breaking structure of fermion-gauge theories is to use low-energy effective Lagrangians \cite{gasser,weinberg} to study the Goldstone modes corresponding to the spontaneously broken global symmetries. Recently, there has been a great deal of progress (for a review see \cite{jac1}) in using such low-energy effective Lagrangians, together with random matrix models, to study the phase structure of nontrivial QCD-like theories with spontaneous symmetry breaking.  An important obstacle, however, is that it is not known how to study these systems at finite baryon density, since the baryon number chemical potential makes
the Euclidean Dirac operator non-Hermitean and the Boltzmann weight
complex. This problem can be overcome, as advocated in \cite{hands,kst,kogut,jac2,jac3}, by considering ``two-color QCD'', for which the fundamental representation of $\SU(2)$ is pseudoreal. This has the consequence that in lattice simulations the Boltzmann weight is real and positive definite,
even at finite (baryon number) chemical potential $\mu$.
Thus, analytic predictions can be quantitatively compared with Monte Carlo
simulations in lattice gauge theory \cite{dagotto,baillie}, provided the lattice regularization respects the relevant flavor symmetry group. Such studies have recently been carried out \cite{kogut2} in 4D QCD with quarks in pseudoreal (and real) representations at
finite $\mu$.

In this paper we study the phase structure of such $
\SU(2)$ QCD-like theories in 3D, at finite baryon density. The possibility of using a real chemical potential still applies in 3D. Our strategy is analogous to the 4D analysis of
Kogut, Stephanov, Toublan, Verbaarschot and Zhitnitsky \cite{kogut}
(denoted KSTVZ hereafter), but the details of the symmetry breaking patterns are completely different in 3D compared to 4D. However, we find that the final answer is almost identical to the 4D case, indicating some degree of universality in the low energy vacuum phase structure. This low energy effective Lagrangian approach assumes that the symmetry breaking occurs, so it is not able to resolve a question such as the existence of a critical number of flavors. Nevertheless, we hope that
our results
may shed some light on the phase structure of the 3D theories, for example in conjunction with lattice analyses.

In 3D, with an {\it even} number
$N_F$ of flavors of massless complex fermions,
denoted by the $N_F/2$ pairs $\psi_f, \chi_f$,
one can predict spontaneous flavor symmetry breaking patterns along similar lines to 4D QCD. In 3D one can introduce of a fermion mass term
\be
{\cal L}_{m}=\sum_{f=1}^{N_F/2} m_{f} (\psib_{f}\psi_{f}-\chib_{f}\chi_{f}).
\label{mass}
\ee
which preserves a discrete symmetry that is a combination of parity and flavor exchange.
The fermion determinant
$
\prod_{f=1}^{N_F/2} \det(-\D^2+m_f^2)
$
for Euclidean space (where the Dirac operator $\D$ is anti-Hermitian)
is positive definite in this case. Therefore one can appeal to the Vafa--Witten theorem \cite{vafa}
and predict that if the flavor symmetry is spontaneously broken in the limit
$m_f\equiv m \to 0$,
the absolute values of the condensates
$\langle\bar{\psi}_f {\psi}_f \rangle$ and
$\langle\bar{\chi}_f {\chi}_f \rangle$ are equal
and their signs are the same as those of respective masses.
That is, the generic situation is that the continuous part of the global symmetry group is broken
according to \cite{pisarski1,poly}
\be
\U(N_F) \rightarrow \U(N_F/2)\times \U(N_F/2)
\label{break2}
\ee
by the quark-antiquark condensate
\be
\sum_{f=1}^{N_F/2}
\left(
\langle\bar{\psi}_f  {\psi}_f \rangle-
\langle\bar{\chi}_f  {\chi}_f \rangle \right),
\label{mcondensate}
\ee
Evidence for such a symmetry breaking pattern has been observed in 3D lattice simulations  \cite{damgaard} with gauge group $\SU(3)$.  This pattern of flavor symmetry breaking can
also be predicted for 3D QCD at large $N_{C}$ using the Coleman--Witten argument \cite{CW}.

The 3D symmetry breaking pattern in (\ref{break2}) is
for quarks in a complex representation of the gauge group,
and is expected to apply to a $\U(1)$ theory and to $\SU(N_C)$ theories with number of colors $N_C\geq 3$. For $\SU(2)$, with fundamental quarks, the symmetry breaking pattern is expected to be different again, due to the pseudoreality of the fundamental representation.  The pseudoreality of the fundamental representation of $\SU(2)$ means that the naive $\U(N_F)$ flavor symmetry is extended to $\USp(2N_F)$, and the continuous part of this global symmetry group is predicted
to break down in 3D as \cite{magnea}
\be
\USp(2N_F) \rightarrow \USp(N_F)\times \USp(N_F).
\label{break1}
\ee
This is different from the generic 3D symmetry breaking pattern in (\ref{break2}), and also is different from the symmetry breaking patterns in 4D theories where the standard flavor symmetry breaking patterns are
\begin{eqnarray}
\U(N_F)_L\times \U(N_F)_R \to \SU(N_F)_V &,& {\rm for}\,\, \SU(N_C\geq 3)\nonumber\\
\U(2N_F)\to \USp(2N_F)   &,& {\rm for}\,\, \SU(2)
\label{4d}
\end{eqnarray}
In 4D the $\U(N_F)$ flavor symmetries are first broken to $\SU(N_F)$ symmetries by the axial anomaly, and then broken by the chiral condensate, with the net breakings as shown in (\ref{4d}).

Physically, the differences between the 3D and 4D cases reflect the differences between
the anomalous discrete symmetries of parity and chirality in 3D and 4D, respectively.
The differences between the gauge groups $\SU(2)$ and $\SU(N_C\geq 3)$ are due to the properties of the representations of these groups. From studies of 4D theories, it has long been appreciated that the $\SU(2)$ theory exhibits exotic types of spontaneous breakdown of global symmetry  \cite{peskin}. Because the representation of the $\SU(2)$ gauge group is either pseudoreal or real, quarks and charge-conjugated antiquarks are combined into an extended flavor multiplet, which is expected to break into its extended vector subgroup. These arguments are actually sensitive only to the representation properties of the gauge group as long as the theory confines. The difference between $\SU(2)$ and $\SU(N_C\geq 3)$  can also be understood in terms of Witten's $\SU(2)$ anomaly in 4D \cite{witten} and its 3D counterpart \cite{redlich}.

In this paper we study the 3D $\SU(2)$ theory, with fundamental quarks, in order to be able to include a real baryon number chemical potential $\mu$. We follow closely the strategy and analysis of KSTVZ, where the 4D SU(2) system with symmetry breaking pattern $\U(2N_F)\to \USp(2N_F)$, as in (\ref{4d}), has been analyzed at finite chemical potential. Since the 3D $\SU(2)$ theory is predicted to have symmetry breaking pattern $\USp(2N_F) \rightarrow \USp(N_F)\times \USp(N_F)$, as in (\ref{break1}), the low-energy degrees of freedom are completely different. Clearly, this means that  the nonlinear $\sigma$ model that describes the Goldstone bosons associated with the flavor symmetry breaking is different from the 4D case studied in KSTVZ.

Since the key feature of the $\SU(2)$ gauge theory that permits a real chemical potential is the existence of the pseudoreal fundamental representation, it is clear that these ideas apply also to  any gauge theory with pseudoreal quarks,
namely $\USp(2N_{C})$
gauge theory with fundamental fermions.
A complementary case with {\em real} quarks, i.e.\
$\SO(N_C)$ gauge theories with fundamental fermions,
and $\SU(N_C\geq 2)$ gauge theories with adjoint fermions,
was also studied at finite chemical potential in 4D in KSTVZ, and yielded
a result universal to the case with pseudoreal quarks.
In 3D the naive $\U(N_F)$ flavor symmetry of the theories with real fermions
is extended to ${\rm O}(2N_F)$, and the predicted flavor symmetry breaking pattern is \cite{magnea2}
\bea
{\rm O}(2N_F) \rightarrow {\rm O}(N_F)\times {\rm O}(N_F).
\label{breakadj}
\eea
Since the pattern of symmetry breaking is different, it is clear that the nonlinear $\sigma$ model that describes the Goldstone bosons associated with the flavor symmetry breaking is different. For the generic 3D flavor breaking in (\ref{break2}) the nonlinear $\sigma$ model has a complex Grassmannian as its target space manifold, while for the pseudoreal fermion case (\ref{break1})  and the real fermion case (\ref{breakadj}) the $\sigma$ model target space is a quaternionic and real Grassmannian, respectively.
Despite these distinctions between pseudoreal and real fermions,
and between the 3D and 4D cases,
we shall show in this paper that the vacuum condensates
and the dispersion laws of the low-lying excitations
share the same functional dependences on the chemical potential.

In section II we briefly review the symmetry breaking pattern in 3D for gauge theories with quarks in pseudoreal and real representations,
at zero and finite chemical potential.
In section III we construct the corresponding low energy effective Lagrangians
by global and local flavor symmetry arguments.
In section IV we analyze the phase structure of the models by finding
the minimum of the effective potential, and compute the condensate
and the baryon density.
In sections V and VI
we expand the potential and kinetic terms,
respectively, of the effective Lagrangian to quadratic order
and derive dispersion laws for the excitations.
In section VII we conclude by providing a group theoretic account for the
observed universality.

\section{global symmetry}
\subsection{Enlarged flavor symmetry}
The fermionic kinetic part of the two-color QCD Lagrangian
with $N_{F}=2n$ flavors of quarks
in Euclidean 3D space
is given by
\be
{\cal L}_{\rm kin}=\psib\D\psi+\chib\D \chi.
\label{Lkin}
\ee
Here $\psi=\psi_{f}^{i}$, $\psid=\psis{}_{f}^{i}$,
$\chi=\chi_{f}^{i}$, $\chid=\chis{}_{f}^{i}$ are independent
two-component spinor fields,
with
the color index $i=1, 2$ and flavor index
$f=1,\ldots,n$ being suppressed.
Pauli matrices denoted as $\sigma_\nu$, with $\nu=1,2,3,$ are employed
to represent the Euclidean Dirac matrix algebra, and those
denoted as $\tau_{\alpha}$, with $\alpha=1,2,3$, are employed
to represent the gauge group algebra.
We choose $x_3$ to be the Euclidean time direction and define
$\psib=\psid \sigma_{3}$, $\chib=\chid \sigma_{3}$.
The Dirac operator is $\D=\sigma_\nu D_\nu$, and
the covariant derivative is
$D_\nu=\partial_\nu+i A_\nu$,
and the gauge field $A_\nu=A_\nu^\alpha \tau_\alpha$
is Hermitian and ${\rm su}(2)$ valued.

We take the parity transformation (P) to be a reflection in the
$x_1$ direction,
\be
x=(x_1, x_2, x_3) \mapsto  x_P=(-x_1, x_2, x_3) .
\ee
Its actions on the  fermion fields are \cite{roman}
\be
\psi(x)  \mapsto  \sigma_1 \psi(x_P)  , \ \
\bar{\psi}(x)  \mapsto  - \bar{\psi}(x_P) \sigma_1 .
\ee
The flavor ${\bf Z}_2$ transformation is an exchange of
$\psi$ and $\chi$,
\be
\psi(x)\leftrightarrow \chi(x),\ \
\bar{\psi}(x)\leftrightarrow \bar{\chi}(x).
\ee
We require the fundamental Lagrangian to be invariant
under the combination of these two transformations,
which we call the (P, ${\bf Z}_2$)--symmetry.
Clearly the kinetic term is (P, ${\bf Z}_2$)--invariant.

Due to the pseudoreality of the $\SU(2)$ Dirac operator,
the Lagrangian (\ref{Lkin}) is invariant under a symmetry group
larger than the apparent $\U(2n)$.
Under the gauge transformation $\psi\to g\psi, \psib\to \psib \gd$,
$g\in \SU(2)$ and
the Lorentz transformation $\psi\to u\psi, \psib\to \psib \ud$,
$u\in \SU(2)$,
a combination
$
\tilde{\psi}=\sigma_{2}\tau_2\psib^T
$
transforms as $\tilde{\psi}\to g \tilde{\psi}$
and $\tilde{\psi}\to u\tilde{\psi}$.
Thus one can put $\psi$, $\chi$,
$\tilde{\psi}$, and $\tilde{\chi}$
into a single flavor $4n$-plet,
\be
\Psi=\left[\ba{cc}\psi\\
\chi\\
\sigma_{2}\tau_2\psib^T\\
\sigma_{2}\tau_2\chib^T
\ea\right] ,
\label{4nplet}
\ee
so that
\be
{\cal L}_{\rm kin}=\frac12 \Psit\sigma_{2}\tau_2\D I \Psi,\ \ \
I= \left[\ba{c|c}& -\openone_{2n}\\ \hline\openone_{2n}& \ea\right].
\ee
As the products of the Pauli matrices $\sigma_2\sigma_\nu$
and  $\tau_2\tau_\alpha$ are symmetric,
so is the operator $\sigma_{2}\tau_2\D$. Accordingly
the matrix acting on flavor indices that appears in the above
fermion bilinear has to be antisymmetric.
In this form the extended flavor symmetry group is
manifestly the unitary symplectic group $\USp(4n)$,
\be
\Psi\to S\Psi, \ \ S^T I S=I,\ \ S^\dagger S=\openone_{4n}.
\ee
The above extension of the flavor symmetry group is analogous to
the two-color QCD in 4D where the conjugated right-handed spinor
$\sigma_{2}\tau_2\psi_R^*$ transforms as the left-handed spinor
$\psi_L$ does
under gauge and Lorentz transformations,
so that the chiral $\SU(N_F)_L\times \SU(N_F)_R$ symmetry
gets extended to $\SU(2N_F)$.

\subsection{Mass term}
The (P, ${\bf Z}_2$)--invariant bare mass term
with degenerate masses,
\be
{\cal L}_{m}=m(\psib\psi-\chib\chi),
\label{massterm}
\ee
can be rewritten as
\be
{\cal L}_{m}=\frac{m}{2} \Psit\sigma_{2}\tau_2 \hat{M} \Psi,\ \
\hat{M}= \left[\ba{cc|cc}
& & -\openone_{n}&\\
& & & \openone_{n}\\
\hline
\openone_{n}&&&\\
& -\openone_{n}&&
\ea\right].
\label{Lm}
\ee
The flavor group $\USp(4n)$ is broken down to $\USp(2n)\times\USp(2n)$
each acting on the (1,3)- and (2,4)-block,
\be
\Psi\to S\Psi,\ \
S=
\left[
\ba{cc|cc}
s_1^{11} & & s_1^{12} &\\
& s_2^{11} & & s_2^{12} \\
\hline
s_1^{21} & & s_1^{22} &\\
& s_2^{21} & & s_2^{22}
\ea  \right],\ \
s_1, s_2 \in \USp(2n) ,
\label{Sp2nxSp2n}
\ee
explicitly by this mass term, or
spontaneously by the quark-antiquark condensate
(\ref{mcondensate})
if formed.
In the latter case, the Goldstone manifold is thus  given
by a quaternionic Grassmannian
$\USp(4n)/(\USp(2n)\times\USp(2n))$. It has $4n^2$ independent
degrees of freedom, twice the three-color case $2n^2$ of the
complex Grassmannian manifold $\U(2n)/(\U(n)\times\U(n))$.
The Grassmannian
$\USp(4n)/(\USp(2n)\times\USp(2n))$ can be canonically parametrized as
\be
\Sigma=S \Sigma_c S^T,\ \ \ \Sigma_c=\hat{M},
\label{Sigmadef}
\ee
where
\be
S(x)=\exp\left(\frac{i \Pi(x)}{2F}\right),\ \ \
\Pi(x)=\pi_a(x) X_a.
\label{S}
\ee
The fields $\pi_a$ are the Goldstone modes, and
the $4n^2$ generators $X_a$ span the subspace
$\usp(4n)-(\usp(2n)\oplus\usp(2n))$.
The construction of the Goldstone manifold (\ref{S})
corresponds to the classification of the generators of
$\usp(4n)$ with respect to a fixed antisymmetric matrix $\Sigma_c$
into $T_k$ and $X_a$;
The $T_k$ generators that span $\usp(2n)\oplus\usp(2n)$
leave $\Sigma_c$ invariant,
\be
\e^{i\phi_k T_k} \Sigma_c (\e^{i\phi_k T_k} )^T=
\Sigma_c ,\ \ \mbox{i.e.}\ \
T_k \Sigma_c =-\Sigma_c T_k^T.
\label{Tk}
\ee
The remaining generators $X_a$ obey the relations
\be
X_a \Sigma_c =\Sigma_c X_a^T,
\ \ \mbox{i.e.}\ \
S\Sigma_c S^T=S^2 \Sigma_c .
\label{Xa}
\ee
The above  partition of the $\usp(4n)$ generators depends on the
matrix $\Sigma_c$. The defining relations (\ref{Xa}) are left
unaltered by rotation of $\Sigma_c$ according to
\be
\Sigma_c\to s \Sigma_c s^T,\ \ s\in \USp(2n)
\ee
accompanied by a simultaneous rotation of the generators by
\be
X_a\to s X_a s^\dagger.
\ee
This means that the set of broken generators, $X_a$, changes
if the matrix $\Sigma_c$ is changed.
The choice of $\Sigma_c=\M^\dagger$  leads to the block representation
of the generators $X_a$ as
\be
\Pi=
\frac12
\left[
\ba{cc|cc}
& P &  & Q \\
P^\dagger &  & Q^T &  \\
\hline
& Q^* &  & -P^* \\
Q^\dagger &  & -P^T &
\ea
\right] ,
\label{Pi}
\ee
where $P$ and $Q$ are $n\times n$ complex matrices,
each having $2n^2$ degrees of freedom.

\subsection{Chemical potential}
The chemical potential that coupled to the baryon number
does not discriminate between the flavors $\psi$ and $\chi$, and is
given by the (P, ${\bf Z}_2$)--invariant form
\be
{\cal L}_{\mu}= - \mu (\psi^\dagger\psi+
\chi^\dagger\chi).
\ee
Without loss of generality we take $\mu\geq 0$.
It can be rewritten in terms of the $4n$-flavored spinor $\Psi$ defined in (\ref{4nplet}),
as
\be
{\cal L}_{\mu}= -
\frac{\mu}{2} \Psi^T i\sigma_1 \tau_2 \hat{C} \Psi, \  \  \
\hat{C}= \left[\ba{c|c}
& \openone_{2n}\\
\hline
\openone_{2n}&
\ea\right]
\ee
This chemical potential term
explicitly breaks
the extended flavor group $\USp(4n)$ down to its
'unextended' $\U(2n)$ subgroup,
\be
\Psi\to S \Psi,
\ \
S=\left[\ba{c|c} U & \\ \hline & U^*\ea\right],\ \ U\in \U(2n).
\label{U2n}
\ee
In the presence of both mass and chemical potential terms,
the surviving global symmetry becomes $\U(n)\times\U(n)$,
\be
\Psi\to S \Psi,
\ \
S=\left[\ba{cc|cc}
u_1 &&&\\
&u_2&& \\
\hline
&& u_1^* &\\
&&&u_2^*
\ea\right],\ \ u_1, u_2\in \U(n),
\label{UnxUn}
\ee
which is the intersection of
$\USp(2n)\times \USp(2n)$ and $\U(2n)$.

\subsection{Diquark source}
\label{VID}
Due to the pseudoreality of the $\SU(2)$ gauge group,
$\tau_2 g \tau_2 = g^*$,
one can write down a gauge invariant bilinear of
two quarks or two antiquarks.
The source term for the
(P, ${\bf Z}_2$)--invariant diquark condensate
\be
{\cal L}_{j}=
j ( \psi^T \sigma_2\tau_2 \chi +
\bar{\psi} \sigma_2\tau_2 \bar{\chi}^T).
\ee
can thus be included (terms like $\psi^T \sigma_2\tau_2 \psi$ vanishes
by anticommutativity).
It can be rewritten in terms of the $4n$-flavored spinor,
\be
{\cal L}_{j}=
\frac{j}{2} \Psi^T \sigma_2 \tau_2 \hat{J} \Psi, \  \J=\left[
\ba{cc|cc}
& \openone_n & & \\
-\openone_n & & &\\
\hline
& & & \openone_n\\
& & -\openone_n&
\ea  \right] .
\ee
This diquark source term and the mass term (\ref{Lm})
belongs to the same multiplet
under the $\USp(4n)$ group, i.e.
there exists a generator $X_2$ of $\USp(4n)$  that rotates $\M$ into $\J$,
\be
\J=\e^{i\frac{\pi}{4} X_2} \M (\e^{i\frac{\pi}{4} X_2})^T, \
X_2=i \left[
\ba{cc|cc}
&& & \openone_n\\
&&\openone_n  &\\
\hline
& - \openone_n&&\\
-\openone_n&&&
\ea  \right] .
\label{X2}
\ee
Namely the sum of mass and diquark source terms
can be comprehensively written as
\bea
{\cal L}_m +{\cal L}_j&=&
\frac{1}{2} \Psi^T \sigma_2 \tau_2 (m\M+j\hat{J}) \Psi
\nn\\
&\equiv&
\frac{m\sec\phi}{2} \Psi^T \sigma_2 \tau_2 \M_\phi \Psi,
\eea
where we have defined the mixing angle $\phi$ by
$\tan \phi={j}/{m}$  and the rotated mass matrix
$\M_\phi=\M \cos\phi+\hat{J}\sin \phi.$
In constructing the effective theory in the following section,
we shall set $j=0$.
The effective Lagrangian
can be translated to the case of nonvanishing $j$
by merely replacing $\M\to \M_\phi$ and $m\to m \sec\phi$.

\section{low-energy effective Lagrangian}
\subsection{Global flavor symmetry}
We consider the effective theory resulting from
the microscopic Lagrangian
\bea
{\cal L}&=&
-\frac{1}{2g^2}\tr F_{\mu\nu}F_{\mu\nu}+
\frac12 \Psit\sigma_{2}\tau_2\D I \Psi\nn\\
&&+\frac{m}{2} \Psit\sigma_{2}\tau_2 \M \Psi-
\frac{\mu}{2} \Psi^T i\sigma_1 \tau_2 \C \Psi,
\label{LQCD}
\eea
which is valid in the
low energy ($\ll\Lambda_{\rm QCD}$) regime
where fundamental particles are confined and
Goldstone bosons dominate.
The kinetic term of the effective Lagrangian
describing the Goldstone modes $\Sigma$, parametrized as in (\ref{Sigmadef}) and (\ref{S}),
should be invariant under the action of the global $\USp(4n)$ group
\be
\Sigma(x)\to s \Sigma(x) s^T, \quad s\in \USp(4n)
\ee
with $s$ in the antisymmetric tensor representation. it should also be
Lorentz invariant, and contain two derivatives.
These requirements uniquely determine its form to be
\be
L_{{\rm kin}}=\frac{F^2}{2} \tr \partial_\nu \Sigma  \partial_\nu \Sigma^\dagger,
\ee
up to a phenomenological ``decay constant'' $F$.
This choice corresponds to normalizing the kinetic term
as defined in (\ref{S}), (\ref{Pi}) to be
the standard form $(1/2)\tr \partial_\nu \Pi \partial_\nu \Pi^\dagger$.
The uniqueness is a consequence of  the pseudo-reality of the field $\Sigma$,
\be
\Sigma^\dagger = I \Sigma I
\label{pr}
\ee
which follows from $\Sigma$ being symplectic, unitary and antisymmetric.
One can systematically generate Skyrme-like higher derivative terms in the effective Lagrangian,
but we do not pursue this issue in this paper as it is not necessary for our applications.

To identify the mass and chemical potential terms in the effective
theory, one uses the well-established strategy
of Gasser and Leutwyler \cite{gasser}.
As the microscopic Lagrangian (\ref{LQCD}) can be made invariant
under the global $\USp(4n)$ transformation
\be
\Psi(x)\to s \Psi(x)
\ee
by promoting the
symmetry-breaking
coupling constant matrices
$m\M$ and $\mu\C$ to field variables that take these
constant condensates in the vacuum
and
assigning the global transformation properties
\be
\M(x)\to s^* \M(x) s^\dagger,\ \ \
\C(x)\to  s^* \C(x) s^\dagger,
\label{MBtransf}
\ee
so should the low-energy effective Lagrangian be similarly invariant.
The collection of local terms allowed by its invariance
will then be reduced to the effective Lagrangian for $\Sigma(x)$
after the replacements $\M(x)\to\M$ and  $\C(x)\to\C$.
To the lowest order in these couplings,
the requirement of invariance determines
the effective mass term to be
\be
L_m=-G\, m\tr (\M \Sigma),
\label{G}
\ee
and the effective chemical potential term
to be
\be
L_\mu= -H \mu^2 \tr (\C \Sigma\C \Sigma),
\label{H}
\ee
The first order term $\tr (\C \Sigma)$ trivially vanishes by symmetry.
Other invariant combinations that appear seemingly different, such as
$\tr (\M^\dagger \Sigma^\dagger)$, or
$ \tr (\C^\dagger \Sigma^\dagger\C^\dagger \Sigma^\dagger),$
$ \tr ( I \C^\dagger \Sigma^\dagger I \C \Sigma)$ and
$ \tr (\C I \Sigma^\dagger \C^\dagger I \Sigma),$ are
equal to $\tr (\M \Sigma)$ or $\tr (\C \Sigma\C \Sigma)$,
respectively, upon substitution
of the vacuum condensates $\M(x)\to\M$ and  $\C(x)\to\C$
and after some algebra using (\ref{pr}).
Namely (\ref{G}) and  (\ref{H}) are already real.

The coefficients $G$ and $H$ in (\ref{G}) and (\ref{H}) are phenomenological constants.
Due to the relationship
\be
\< \psib \psi -\chib\chi \> = -\frac{1}{V}\frac{\partial}{\partial m}Z(m,\mu),
\label{GZ}
\ee
and the  vacuum alignment of the field
$\Sigma(x)\to\Sigma_c=\M^\dagger$ for $m\searrow 0, \mu\to 0$,
[which will be shown below -- see eq.(\ref{psibpsi})],
the constant $G$  is equal to the
quark-antiquark condensate
at $m, \mu \to0$,
\be
G=\frac{1}{4n}\,\lim_{m\searrow 0}\left.\< \psib \psi -\chib\chi \>\right|_{\mu=0}.
\ee

\subsection{Local flavor symmetry}
Unlike the quark-antiquark condensate
$G$, the phenomenological parameter $H$ is not independent of
the decay constant $F$, but
is related to it by virtue of
the local flavor symmetry \cite{kst}.
To lift the global $\USp(4n)$ symmetry to a local one,
we combine ${\cal L}_{{\rm kin}}$ and
${\cal L}_{\mu}$ into a covariant derivative form
acting on the extended quark multiplet $\Psi$,
\bea
{\cal L}_{{\rm kin}} + {\cal L}_{\mu}&=&
\frac12 \Psit\sigma_{2}\tau_2
\sigma_\nu  I (
D_\nu  - \mu B_\nu ) \Psi,
\nn\\
B_\nu&=&\B \delta_{\nu,3}  ,\ \
\B=\C I=
\left[
\ba{c|c}
\openone_{2n} & \\
\hline
& - \openone_{2n}
\ea
\right] .
\label{B}
\eea
This Lagrangian is invariant under  the {\em local} $\USp(4n)$ symmetry
\be
\Psi(x)\to s(x) \Psi(x),\ \ s(x)\in \USp(4n),
\ee
when $B_\nu$ is promoted to a local field and is
assigned the inhomogeneous transformation property of
a flavor gauge field,
\be
B_\nu(x) \to  s(x) B_\nu(x) s(x)^\dagger
+\frac{1}{\mu} \left(\partial_\nu s(x) \right) s(x)^\dagger ,
\label{flavorgauge}
\ee
rather than the homogeneous global transformation (\ref{MBtransf}).
This local symmetry should also be respected by the effective
Lagrangian, whose field variable transforms as
\be
\Sigma(x)\to s(x)\Sigma(x)s(x)^T.
\ee
To cancel the extra two terms that appear in the
transformation of the derivative of the field,
\be
\partial_\nu \Sigma\to s(\partial_\nu\Sigma) s^T+
(\partial_\nu s)\Sigma s^T +
s \Sigma( \partial_\nu s^T) ,
\ee
one introduces flavor-covariant derivatives for the fields
in the antisymmetric tensor representations
\bea
\nabla_\nu \Sigma&=&
\partial_\nu \Sigma-
\mu(\Sigma B_\nu^\dagger + B_\nu^\dagger \Sigma)\nn\\
\nabla_\nu \Sigma^\dagger&=&
\partial_\nu \Sigma^\dagger+
\mu(\Sigma^\dagger  B_\nu + B_\nu  \Sigma^\dagger).
\eea
They transform covariantly
\be
\nabla_\nu \Sigma\to s(\nabla_\nu \Sigma)s^T,\ \
\nabla_\nu \Sigma^\dagger\to s^*(\nabla_\nu \Sigma^\dagger)s^\dagger,
\ee
due to the gauge transformation property
(\ref{flavorgauge}), while preserving the antisymmetry.
Accordingly, the total effective Lagrangian reads
\bea
L&=&L_{{\rm kin}}+L_\mu+L_m
\label{Leff}\\
&=&\frac{F^2}{2}
\tr \nabla_\nu \Sigma\nabla_\nu \Sigma^\dagger -G m\tr (\M \Sigma)
\nn\\
&=&
\frac{F^2}{2} \tr \partial_\nu \Sigma  \partial_\nu \Sigma^\dagger
+
{F^2 \mu} \,{\rm tr}
\left(
(\Sigma^\dagger \B + \B \Sigma^\dagger  ) \partial_3 \Sigma
\right)
\nn\\
&-&
{F^2\mu^2}
\left( \tr(\B\Sigma \B\Sigma^\dagger)+4n\right)-{F^2 m_\pi^2}{} \tr (\M \Sigma) \nn
\eea
after partial integration on the linear derivative term.
We observe that just as in \cite{kst},
the parameter $H$ is not independent, but equal to
$F^2$ .
In the last line of (\ref{Leff}), we again used the pseudoreality of $\Sigma$ (\ref{pr}), and
also
replaced $G$ by the mass of the
pseudo-Goldstone bosons $\pi_a(x)$ in (\ref{S}) by
the Gell-Mann--Oakes--Renner relation
\be
m^2_\pi = \frac{Gm}{F^2}
\ee
that results by expanding the fluctuation of the field
$\Sigma(x)$ to second order (see the next section).

\section{vacuum alignment and condensates}
The static part of the effective Lagrangian (\ref{Leff}),
i.e.\ the effective potential,
determines the vacuum alignment of the field $\Sigma$.
As in \cite{kst}, we introduce a dimensionless phenomenological parameter
$\xi=2\mu/m_\pi$ to represent the chemical potential, and write
\be
L_{\rm st}(\Sigma)={F^2 m_\pi^2}{}\left(
-\tr (\M \Sigma)
-\frac{\xi^2}{4} \tr( \B\Sigma\B\Sigma^\dagger )- n\xi^2\right).
\label{Veff}
\ee
The above two terms compete for the direction of the condensate
which we denote by $\bar{\Sigma}$.

To determine the vacuum condensate,
we write $\Sigma$ in a block form (with each sub-block being $2n\times 2n=N_F\times N_F$):
\be
\Sigma=
\left[
\ba{c|c}
\Sigma_{11} & \Sigma_{12}\\
\hline
-\Sigma_{12}^T & \Sigma_{22}
\ea
\right] .
\ee
These component matrices also satisfy unitarity constraints
\bea
&&\Sigma_{11}\Sigma_{11}^\dagger+
\Sigma_{12}\Sigma_{12}^\dagger=\openone_{2n}, \nn\\
&&\Sigma_{12}^T\Sigma_{12}^*+
\Sigma_{22}\Sigma_{22}^\dagger=\openone_{2n}, \nn\\
&&\Sigma_{11}\Sigma_{12}^*=
\Sigma_{12}\Sigma_{22}^\dagger ,
\label{unitarity}
\eea
as well as constraints arising from the fact that $\Sigma$ is symplectic and antisymmetric.
Using solely the unitarity constraints and the reality of
$\tr\M\Sigma$,
one can express
$L_{\rm st}$ entirely in terms of $\Sigma_{12}$,
\bea
L_{\rm st}(\Sigma)&=&F^2 m_\pi^2 \left[
\xi^2 \tr \Sigma_{12}\Sigma_{12}^\dagger
-\tr \Gamma \Sigma_{12} \right.\nn\\
&&\hskip 2cm \left.
-\tr (\Gamma \Sigma_{12})^\dagger
-2n\xi^2\right]
\nn\\
&=&\hskip -5pt
F^2m_\pi^2 \left[\xi^2
\tr( \Sigma_{12}\Gamma-\xi^{-2}\openone_{2n})
( \Sigma_{12}\Gamma-\xi^{-2}\openone_{2n})^\dagger\right.
\nn\\
&&\hskip 2cm\left. -2n (\xi^2+\xi^{-2})\right],
\label{Trsquare}
\eea
where
\be
\Gamma=
\left[
\ba{cc}
\openone_{n} & \\
& -\openone_{n}
\ea
\right] .
\ee

If we ignore the constraints on $\Sigma_{12}$ for a moment,
the trace in (\ref{Trsquare}) can be viewed from the distance
between two points $\Sigma_{12}$ and $\xi^{-2}\openone_{2n}$
in the $2n^2$-dimensional vector space of real and imaginary parts of
the matrix elements.
For $\xi>1$ the absolute minimum is achieved when
$\Sigma_{12}=\xi^{-2}\Gamma$.
The unitary, symplectic, and antisymmetry constraints on $\Sigma$ then determine
the other sub-block matrices to be, up to the $\U(n)$ residual degree of freedom
$u\in\U(n)$,
\bea
&&\Sigma_{11}=\sqrt{1-\xi^{-4}}
\left[
\ba{cc}
& u \\
-u^T &
\ea
\right] ,\nn \\
&&\Sigma_{22}=\sqrt{1-\xi^{-4}}
\left[
\ba{cc}
& u^* \\
-u^\dagger &
\ea
\right]  .
\eea

When $\xi<1$, the unitarity constraints (\ref{unitarity}) demand that
\be
\tr( \Sigma_{12}\Gamma)(\Sigma_{12}\Gamma)^\dagger \leq 2n,
\ee
and the point  $\xi^{-2}\openone_{2n}$ lies outside of this region.
The point closest to   $\xi^{-2}\openone_{2n}$
within this $2n^2$-dimensional ball is clearly at
$\Sigma_{12}\Gamma=\openone_{2n}$.
Then the unitarity constraints  (\ref{unitarity}) can  be satisfied
only by $\Sigma_{11}=\Sigma_{22}=0$.
To summarize, the condensate that gives the global minimum
of the static effective Lagrangian is
\be
\bar{\Sigma}=
\M^\dagger \cos \alpha +
\left[
\ba{cc|cc}
& u & & \\
-u^T & & &\\
\hline
& & & u^*\\
& & -u^\dagger &
\ea
\right] \sin\alpha,
\label{barSigma}
\ee
where $u\in \U(n)$, and we have defined
\be
\cos\alpha=\min ( 1,  \xi^{-2} ) .
\label{cosr}
\ee
Note that the first term of the condensate (\ref{barSigma}) does not carry
baryon number, while the second term does.

The condensate is a non-analytic function of $\xi=2\mu/m$.
In the regime $\xi>1$,  the vacuum condensate
has the $\U(n)$ degeneracy corresponding to
$n^2$ true Goldstone modes.
This change of massless modes indicates
a second-order phase transition at $\xi=1$.
The static effective Lagrangian is
\bea
L_{\rm st}(\bar{\Sigma})
&=&
-4nF^2 m^2_\pi \times\left\{
\ba{ll}
1 &\ \ (\xi<1)\\
(\xi^{2} +\xi^{-2})/2 &\ \ (\xi>1)
\ea
\right . \nn\\
&=&
\left\{
\ba{ll}
-4n G m  &\ (\mu<m_\pi /2)\\
-n \left(
{8F^2 \mu^2}{} +\frac{G^2 m^2} {2F^2 \mu^2}
\right)  &\ (\mu>m_\pi /2)
\ea
\right.
\label{Evac}
\eea
The second $\xi$-derivative of the $L_{\rm st}$
is indeed discontinuous at $\xi=1$.

The quark-antiquark condensate
$\< \psib \psi - \chib \chi \>$ and
the baryon density
$\< \psid \psi + \chid \chi \>$ directly follow from (\ref{Evac}).
Upon differentiation with respect to $m$ and $\mu$, respectively, we find
\bea
\< \psib \psi - \chib \chi \> &=&
\left\{
\ba{ll}
4nG &\ \ (\xi<1)\\
4nG \xi^{-2} &\ \ (\xi>1)
\ea
\right . ,
\label{psibpsi}\\
\< \psid \psi + \chid \chi \> &=&
\left\{
\ba{ll}
0 &\ (\xi<1)\\
16nF^2 \mu  (1-\xi^{-4}) &\ (\xi>1)
\ea
\right. .
\label{psidpsi}
\eea
These condensates are identical (with the replacement $2n\to N_f$) to the corresponding condensates in the 4D $SU(2)$ case, with fundamental quarks, studied in \cite{kogut} -- see Table 3 in \cite{kogut}. This follows from the fact that the vacuum energy has the same functional form in the 3D and 4D cases. This is nontrivial from the microscopic point of view since the symmetries involved are completely different. Rather, this universality arises from the mean field treatment of the low-energy effective theory, as we discuss further in the conclusions.

\section{curvature at the minimum}
In this section we expand the static effective Lagrangian
around the vacuum condensate (\ref{barSigma}) to
second order. In the diquark condensation phase $\xi>1$,
we arbitrarily fix the $\U(n)$ degeneracy by $u=\openone_n$,
\be
\bar{\Sigma}_\alpha=\M^\dagger \cos \alpha + \J \sin \alpha,
\label{bSigma}
\ee
with $\alpha$ given by (\ref{cosr}).
At small chemical potential in the range
$0\leq\xi<1$ ($\alpha=0$) this condensate aligns to $\M^\dagger$,
which is preserved by  the $\USp(2n)\times\USp(2n)$
subgroup acting on the (1,3) and (2,4) blocks :
$\Sigma\to S\Sigma S^T$,
with $S$ given by (\ref{Sp2nxSp2n}).
Namely for $0< \xi<1$ the mass term explicitly breaks
this symmetry down to $\U(n)\times\U(n)$ given by
(\ref{UnxUn}), also
acting on the (1,3) and (2,4) blocks.
On the other hand, in the massless limit
$\xi=\infty$ this condensate aligns to $\J$,
and the residual global symmetry becomes the $\USp(2n)$ subgroup,
\be
\Sigma\to
\left[
\ba{c|c}
s& \\
\hline
& s^*
\ea  \right]
\Sigma
\left[
\ba{c|c}
s^T & \\
\hline
& s^\dagger
\ea  \right],\  \
s \in \USp(2n),
\ee
of the flavor $\USp(4n)$ group.
In the intermediate region $1<\xi<\infty$ the residual symmetry is the
$\U(n)$ subgroup given by (\ref{UnxUn}) with $u_2=u_1^*$.
All excitations will be classified according to the representations of
these residual symmetry groups.

\subsection{Normal phase}
When $\xi<1$ the vacuum orientation of the condensate
does not depend on $\xi$ and is given by $\bar{\Sigma}=\M^\dagger$.
Expanding $\Sigma$ around $\M^\dagger$
using the Goldstone field defined in (\ref{S}) according to
\be
\Sigma=S \M^\dagger S^T=S^2 \M^\dagger=
\( 1+\frac{i\Pi}{F}-\frac{\Pi^2}{2F^2}+\cdots \)\M^\dagger,
\ee
we find
\be
L_{{\rm st}}(\Sigma)=
L_{{\rm st}}(\bar{\Sigma}_0)+
\frac{m^2_\pi}{2} \left[
\tr \Pi^2 +\frac{\xi^2}{4} \tr [\B,\Pi]^2 \right]
+\cdots,
\ee
where the ellipsis denotes terms of higher order in $\Pi$.
Substituting the parametrization (\ref{Pi}), we obtain
\bea
L_{{\rm st}}(\Sigma)&=&
L_{{\rm st}}(\bar{\Sigma}_0)\\
&+&
\frac{m_\pi^2}{2}  \left[
\tr PP^\dagger + (1-\xi^2)\tr QQ^\dagger
\right]
+\cdots ,\nn \\
L_{{\rm st}}(\bar{\Sigma}_0)&=&
-4n F^2 m_\pi^2. \label{Evacn}
\eea
This confirms that the configuration (\ref{bSigma}) is indeed a minimum.
We also see that there are no true Goldstone modes for $\xi<1$.
At $\xi=1$ the curvature of the diquark modes $Q$ vanish,
which signals a second order phase transition and the
diquark condensation.

Below we list the representations of the residual symmetry group
$\U(n)\times \U(n)$ to which the fields belong
by listing the representations' Young tableaux,
\bea
P\ \ &:&\ \ \(\ \bx\ , \ \bxs\ \)
\nn\\
P^\dagger\ \ &:&\ \  \(\ \bxs\ , \ \bx\ \)
\nn\\
Q\ \ &:&\ \ \(\ \bx\ , \ \bx\ \)
\nn\\
Q^\dagger\ \ &:&\ \ \(\ \bxs\ , \ \bxs\ \) ,
\eea
where $\bx$ and $\bxs$ stand for fundamental and
conjugated fundamental representations, respectively.
The dimensions of these representations are all equal to $n^2$.
At vanishing chemical potential,
$\xi=0$, these four fields combine to become
a $4n^2$--dimensional
multiplet $\(\ \bx\ , \ \bx\ \)$ of $\USp(2n)\times\USp(2n)$.

\subsection{Diquark condensation phase}
When $\xi>1$ the configuration (\ref{bSigma})
begins to rotate according to $\cos \alpha = \xi^{-2}$.
This rotation can also be written as
\be
\bar{\Sigma}_\alpha= s_\alpha \M^\dagger s_\alpha^T = s_\alpha^2 \M^\dagger,\ \
s_\alpha = \e^{i \frac{\alpha}{2} X_2},
\ee
where
$X_2$ is the generator, defined in (\ref{X2}), that rotates $\M$ into $\J$ .
We could parametrize the fluctuation
around the vacuum $\bar{\Sigma}_\alpha$ as
\be
\Sigma=S_\alpha \bar{\Sigma}_\alpha S_\alpha^T
=S_\alpha s_\alpha \M^\dagger   s_\alpha^T S_\alpha^T ,
\label{SsMsS}
\ee
where $S_\alpha$ are symplectic matrices generated by rotated generator
$s_\alpha X_a s_\alpha^\dagger$, instead of $X_a$.
However,
we employ an alternative parametrization of the fluctuation,
\be
\Sigma=s_\alpha S \M^\dagger  S^T  s_\alpha^T,
\label{sSMSs}
\ee
where $S$ are generated by unrotated generators $X_a$,
in order for the meson mass matrix to be diagonal subsequently.
We substitute this parametrization into the static effective Lagrangian
(\ref{Veff})
and obtain
\bea
&&L_{\rm st}(\Sigma)={F^2 m_\pi^2}\left[-
\tr ( \M^\dagger s_\alpha^T \M s_\alpha S^2)  \right.
\label{MKBL}\\
&&
\left. -\frac{\xi^2}{4} \tr(\M^\dagger s_\alpha^T \B s^*_\alpha \M
(S^\dagger)^2 s^\dagger_\alpha \B s_\alpha  S^2)
-n\xi^2\right] .
\nn
\eea
Substituting the parametrizations (\ref{S}) and (\ref{Pi}) into (\ref{MKBL}),
and expanding to second order in
$P$ and $Q$, we find
\bea
L_{{\rm st}}(\Sigma)&=&
L_{{\rm st}}(\bar{\Sigma}_\alpha)
+
\frac{m^2_\pi}{2}\left[ \xi^2 \tr P_S P_S^\dagger+\xi^{-2} \tr P_A P_A^\dagger
\right.\nn\\
& &+\left.(\xi^2-\xi^{-2}) \tr Q_I^2  \right]
+\cdots .
\label{Veffdq}\\
L_{{\rm st}}(\bar{\Sigma}_\alpha)&=&
-2n(\xi^2+\xi^{-2}) F^2 m_\pi^2.
\label{Evacdq}
\eea
Here we have defined $P_S, P_A, Q_R, Q_I$ as
\bea
P_S=\frac{P+P^T}{2}=P_S^T
\ \ &:&\ \
\ba{|c|c|}\hline\ \ &\ \ \\ \hline\ea \ ,\
\ba{|c|c|}\hline\,\ast \, &\, \ast \, \\ \hline\ea
\nn\\
P_A=\frac{P-P^T}{2}=-P_A^T
\ \ &:&\ \
\ba{|c|}\hline\ \ \\ \hline \ \ \\ \hline\ea \ ,\
\ba{|c|}\hline\,\ast \, \\ \hline \,\ast \, \\ \hline\ea
\nn\\
Q_R=\frac{Q+Q^\dagger}{2}=Q_R^\dagger
\ \ &:&\ \
\ba{|c|c|}\hline\,\ast \, &\ \ \\ \hline\ea
\nn\\
Q_I=\frac{Q-Q^\dagger}{2i}=Q_I^\dagger\, .
\ \ &:&\ \
\ba{|c|c|}\hline\,\ast \, &\ \ \\ \hline\ea
\eea
These projections are orthogonal,
$\tr P_S P_A=\Im m\tr Q_R Q_I =0$.
The representations of the residual symmetry group
$\U(n)$ to which the fields belong are denoted
by the Young tableaux.
The dimensions of  these representations are
$n^2+n$, $n^2-n$, $n^2$ and $n^2$, respectively.

{}From (\ref{Veffdq}) we can read off the curvatures of the different
multiplets of the pseudo-Goldstone modes,
and again confirm the local minimality of the
configuration (\ref{bSigma}).
We see that there are $n^2$ true flat directions, $Q_R$,
which describe massless Goldstone modes.
By contrasting this fact with the solution (\ref{barSigma})
at $\xi\to\infty$,
the true Goldstone fields are identified with
the $ \U(n)$ phases of the diquark condensate.
As we shall see in the next section,
two modes
$Q_R$ and $Q_I$ that belong to the same representation
are mixed by
the linear derivative terms in the effective Lagrangian,
and thus the actual true Goldstone
excitations are certain linear combinations of
$Q_R$ and $Q_I$.

\section{mass spectrum}
In order to determine the spectrum of low-lying excitations
we must take into account the derivative terms in the
effective Lagrangian (\ref{Leff}).
The dispersion laws
of the system at finite chemical potential
do not take the simple  Lorenz invariant form
$E^2={\bf p}^2+M^2$.
We shall evaluate the pole mass (rest energy)
that is the value $E=i p_3$ of the pole  of the
propagator at ${\bf p}=0$.

\subsection{Normal phase}
Expanding the derivative terms  in the effective
Lagrangian (\ref{Leff}) in the same way as we expanded the
potential part in the previous section, we obtain
\bea
L&=&L_{\rm st}(\bar{\Sigma}_\alpha)+
\frac12\tr  \partial_\nu P^\dagger  \partial_\nu P
+\frac{m^2_\pi}{2} \tr  P^\dagger   P
\nn
\\
&+&\frac12 \tr \partial_\nu Q^\dagger  \partial_\nu Q
- 2\mu \tr Q^\dagger  \partial_3 Q
+\Bigl( \frac{m_\pi^2}{2}-2\mu^2 \Bigr) \tr  Q^\dagger   Q
\nn\\
&+&\cdots .
\label{Lquad}
\eea
After using the Fourier decomposition of $\Pi$,
\be
\Pi(x)=\sum\nolimits_{p} \Pi_p \,\e^{-ipx},
\ \ p=({\bf p}, -iE)
\ee
the equations of motion for $P, P^\dagger, Q, Q^\dagger$
that follow from (\ref{Lquad}) lead to the dispersion law
\be
(E+b \mu)^2= {\bf p}^2 + m_\pi^2 .
\ee
Here $b$ is the baryon charge of the corresponding excitation,
assigned as :
$b=0$  for $P$ and $P^\dagger$;
$b=2$ for $Q$;  and
$b=-2$ for $Q^\dagger$.
Accordingly, the pole masses for these excitations are
\bea
m_\pi &{\rm for}& P\,\, {\rm and}\,\, P^\dagger\nn\\
m_\pi-2\mu &{\rm for}& Q\nn\\
m_\pi+2\mu &{\rm for}& Q^\dagger
\eea

\subsection{Diquark condensation phase}
In this phase the ground state changes and we need to
expand around the rotated value of the condensate
$\bar{\Sigma}_\alpha$, as in (\ref{sSMSs}).
To second order in the fluctuation $\Pi$ we obtain
\bea
L&=&L_{\rm st}(\bar{\Sigma}_0)
+\frac12\tr  \partial_\nu P_S^\dagger  \partial_\nu P_S
+\frac12\tr  \partial_\nu P_A^\dagger  \partial_\nu P_A
\nn\\
&+&\frac12 \tr (\partial_\nu Q_R)^2
+\frac12 \tr (\partial_\nu Q_I)^2
\nn\\
&+&2i \mu\xi^{-2} \tr
(Q_R^\dagger \partial_3 Q_I
-Q_I^\dagger \partial_3 Q_R)
\nn\\
&+&
\frac{m^2_\pi}{2}\left[
 \xi^2 \tr P_S P_S^\dagger
+\xi^{-2} \tr P_A P_A^\dagger
+(\xi^2-\xi^{-2}) \tr Q_I^2  \right]
\nn\\
&+&\cdots .
\label{Lquaddq}
\eea
The linear derivative term contains only the $Q$ fields,
which have a nonzero baryon charge, and the dispersion law for the $P$ fields
remain Lorentz invariant in form:
\bea
P_S \ &:&\ E^2= {\bf p}^2 + m_\pi^2\xi^2 ,\nn\\
P_A \ &:&\ E^2= {\bf p}^2 + m_\pi^2\xi^{-2} .
\eea
On the other hand,
the dispersion laws for the $Q$ fields are determined by
the secular equation obtained by substitution of the Fourier
decomposition of $Q_R$ and  $Q_I$,
\be
\det
\left[
\ba{cc}
[E^2-{\bf p}^2]  & 4iE\,\mu \xi^{-2} \\
-4iE\,\mu \xi^{-2} &[E^2-{\bf p}^2  -m^2_\pi(\xi^2-\xi^{-2})]
\ea
\right]
=0.
\label{secular}
\ee
We see that $Q_R$ and $Q_I$ are mixed by the linear derivative
term unless $\cos \alpha=0$, i.e.\ in the massless limit.
Due to the form of (\ref{secular}), for any $\alpha$ there is
a solution for which $E({\rm p}=0)=0$, and the true Goldstone mode is
denoted by $\Q$. In the massless limit it is entirely $Q_R$.
The other massive solution of the secular equation,
which is a linear combination of $Q_R$ and $Q_I$ orthogonal
to $\Q$, is denoted by $\Q^\dagger$.
The dispersion laws for these fields are given by
\bea
\Q \ \ :\ \ E^2&=& {\bf p}^2 + 2\mu^2 (1+3\xi^{-4})
\nn\\
&-&2\mu\sqrt{\mu^2(1+3\xi^{-4})^2 +4 {\bf p}^2\xi^{-2} },
\nn\\
\Q^\dagger \ \ :\ \ E^2&=& {\bf p}^2 + 2\mu^2 (1+3\xi^{-4})
\nn\\
&+&2\mu\sqrt{\mu^2(1+3\xi^{-4})^2 +4 {\bf p}^2\xi^{-2} }.
\eea
The pole masses of these four excitations, i.e.\ $E({\rm p}=0)$, are
then given by
\bea
2\mu &{\rm for}& P_S\nn\\
\frac{m^2_\pi}{2\mu}&{\rm for}&  P_A\nn\\
0&{\rm for}& \Q\nn\\
2\mu\sqrt{1+3\left( \frac{m_\pi}{2\mu}\right)^4}&{\rm for}&\Q^\dagger
\eea
These dispersion laws and pole masses are
identical to their 4D counterparts: see Eq.(86) of KSTVZ.
This is a remarkable manifestation of the
universality of the low-energy
effective theories governing quasi-Goldstone bosons.

We finally note that
in the massless limit
$P_S$ and $\Q^\dagger= Q_I$ combine to become
a $(2n^2+n)$--dimensional multiplet
$\ba{|c|c|}\hline\ \ &\ \ \\ \hline\ea$ of $\USp(2n)$, and
$P_A$ and $\Q= Q_R$ combine to become
a $(2n^2-n)$--dimensional multiplet
$\ba{|c|}\hline\ \ \\ \hline \ \ \\ \hline\ea$ of $\USp(2n)$.

\section{conclusion}
In this paper we have used the
effective field theory method at finite baryon density to
investigate the phase structure and the
excitation spectrum of the three dimensional parity invariant
SU(2) QCD with fundamental quarks. Since the quarks are in a pseudoreal representation we can introduce a real chemical potential for baryon number, as was done for 4D in \cite{kst,kogut}.
Our main result is that even though the
symmetry breaking groups are different in 3D and 4D, the final phase structure, condensates and dispersion relations are the same in 3D and 4D.
Furthermore, it is straightforward to
generalize this to the case with adjoint fermions
(and any $\SU(N_C)$ gauge group);
at the level of fundamental Lagrangian
one merely needs to ignore the color $\tau_2$ matrices
altogether.
This leads to
replacing the $\USp$
groups that appear in the global symmetry
arguments by the ${\rm O}$ groups of the same dimension
due to the change of the constant matrices,
\bea
I&=& -\left[\ba{c|c}& \openone_{2n}\\ \hline \openone_{2n}& \ea\right], \,
\hat{M}= \left[\ba{cc|cc}
& & -\openone_{n}&\\
& & & \openone_{n}\\
\hline
-\openone_{n}&&&\\
& \openone_{n}&&
\ea\right],
\nn\\
\hat{C}&=& \left[\ba{c|c}
& \openone_{2n} \\
\hline
-\openone_{2n}&
\ea\right], \,
\hat{J}= \left[\ba{cc|cc}
& \openone_{n}&&\\
\openone_{n}&&&\\
\hline
&&&-\openone_{n}\\
&& -\openone_{n}&
\ea\right]
\eea
(the matrix $\B=\C I$ is left unchanged from (\ref{B})
except for an irrelevant flip of the sign),
and accordingly to
interchanging the symmetric and antisymmetric properties
of the field $\Sigma(x)$ and the representations of various
excitations.
Thus we have shown
that the low-energy behavior of  (pseudo-)real gauge theories
at finite baryon density
is universal beyond the dimensionality of spacetime
and the representation of the quarks.
The  individuality of each case resides solely in the number of modes belonging to
a multiplet with common mass.

To compute the
diquark condensate
$\< \psi^T \sigma_2\tau_2 \chi -
\psi^\dagger \sigma_2\tau_2 \chi^* \>$,
one should include its source term in the effective Lagrangian
by replacing the matrix
$\M$ by $\M_\phi$ and $m$ by $m\sec\phi$,
as mentioned at the end of sect.\ref{VID}.
This modification of the mass term
smears the abrupt change of the chemical potential dependence
of the vacuum condensate at $\xi=1$,
and turns the second-order phase transition into a crossover, just as in the 4D case \cite{kogut}.
This is a straightforward exercise and we do not record the details here. The final results for the
chemical potential dependences of the
quark-antiquark condensate, diquark condensate, and
baryon density as well as the dispersion laws for the excitations
are identical to the 4D case.

The universality becomes most transparent by expressing
the whole argument in terms of  the generic symmetric spaces
\cite{gilmore}
that arise when a global symmetry group is broken.
Let $G$ be a compact Lie group and ${\mathcal{G}}$
its Lie algebra $G=\exp{\mathcal{G}}$.
For $G$ to be identified as a global symmetry of a gauge theory
with (pseudo-)real quarks, its dimension is set to be even.
An involutive automorphism $\sigma$ of $\mathcal{G}$ is defined as
\be
\sigma: \cG\to \cG,\ X\mapsto \sigma(X),
\ \sigma^2=1 .
\ee
An involution $\sigma$ naturally introduces a decomposition (known as  the Cartan decomposition) of ${\mathcal{G}}$
with respect to $\sigma$.  This decomposition splits ${\mathcal{G}}$ into
a subalgebra with eigenvalue $+1$, and a subspace orthogonal to it
with eigenvalue $-1$,
\be
\mathcal{G}=\mathcal{K}\oplus \mathcal{P},\ \
\sigma(X) = \pm X\ \mbox{if}\  X \in\left\{ {\mathcal{K} \atop \mathcal{P}}\right. .
\ee
Below we will identify certain choices for this involution with mass and chemical potential terms.

Denote by $K$ the Lie subgroup generated by $\mathcal{K}$,
$K=\exp\mathcal{K}$.
The Riemannian symmetric space $G/K$
is isomorphic to the exponential map of
$\mathcal{P}$, $G/K\simeq\exp\mathcal{P}$
through the Cartan mapping $\mathcal{C}$,
\be
\mathcal{C}(g)=
g \,\sigma(g)^{-1} .
\ee
Now we choose an involution of the form
\be
\sigma(g)=\M^{-1} g^* \M, \quad
\M^* \M=\pm 1,\quad
\M^\dagger \M=1,
\ee
which introduces a Cartan decomposition $\cG=\cK\oplus\cP,$
and specifies the unbroken vector subgroup $K$, determined by the dynamics.
This involution will be associated with the mass term.
The low energy effective theory for the Goldstone modes is a nonlinear $\sigma$
model on $G/K$; that is, it is a field theory
whose field variable is parametrized by $\mathcal{C}(g)$ for $g\in G$.
The potential of the nonlinear sigma model
should also be invariant under the left action of
$K$ on $G$, $g\mapsto k g$, which
acts adjointly on $\mathcal{C}(g)$ by
$\mathcal{C}(g)\mapsto k\mathcal{C}(g)k^{-1}.$
Such an invariant quantity is spanned by
$\{ \tr \mathcal{C}(g)^\nu\}_{\nu\in {\bf Z}}$. We define the mass term in the
effective field theory to be the combination that is real and of lowest order, i.e.\
$\Re{\rm e} \tr  \mathcal{C}(g)$.

Next we fix another involution of the form
\be
\tilde{\sigma}(g)=\B^{-1} g \B,\quad  \B^2=1.
\ee
which introduces another Cartan decomposition
$\cG=\tilde{\cK}\oplus \tilde{\cP}$, with associated Cartan mapping
$\tilde{\mathcal{C}}(g)=g \tilde{\sigma}(g)^{-1}$.
For this involution $\tilde{\sigma}$ to be identified with
the baryon charge, we take
$\B={\rm diag}(1,\ldots,1, -1,\ldots,-1).$
The corresponding symmetry breaking term  in the effective
Lagrangian must be expressed in terms of the field variable $\mathcal{C}(g)$, and must be invariant only under the left action of
$K\cap\tilde{K}$ on $G$. These terms are spanned by
$\{\tr \tilde{\mathcal{C}}( \mathcal{C}(g)^\nu)^\lambda
\}_{\nu, \lambda\in {\bf Z}}$. We define the chemical potential term to be the combination that is real and of lowest nontrivial order, i.e.\
$\tr  \tilde{\mathcal{C}}(\mathcal{C}(g))$.

Thus, the static part of the effective Lagrangian is
\be
L_{\rm st}=-\Re{\rm e}\tr \mathcal{C}(g)+\frac{\xi^2}{4}
\tr  \tilde{\mathcal{C}}(\mathcal{C}(g))
\ee
In order to find the extremum of  $L_{\rm st}$, we decompose $\mathcal{C}(g)$
into four blocks according to those of $\B$,
\be
\mathcal{C}(g)=
\left[
\ba{c|c}
\mathcal{C}_{11} & \mathcal{C}_{12} \\
\hline
\mathcal{C}_{21} & \mathcal{C}_{22}
\ea
\right] .
\ee
Using the unitarity constraints on $\mathcal{C}_{ij}$,
$L_{\rm st}$ now reads
\bea
L_{\rm st}&=&
\frac{\xi^2}{2} \tr
\left( \mathcal{C}_{11} - \xi^{-2} \right)
\left( \mathcal{C}_{11} - \xi^{-2}\right)^\dagger
\nn\\
&+&
\frac{\xi^2}{2} \tr
\left( \mathcal{C}_{22} - \xi^{-2}\right)
\left( \mathcal{C}_{22} -  \xi^{-2}\right)^\dagger
+{\rm const.}
\eea
This factorized form covers the various 4D cases treated in KSTVZ \cite{kogut}, as well as the 3D cases considered in this paper. The important point for the universality of the form of this static effective Lagrangian is that one does not need to specify whether the original global symmetry group is $\SU(2N_F)$ or $\USp(2N_F)$ or ${\rm O}(2N_F)$. The two-phase structure is manifest from this form of $L_{\rm st}$, as shown in \cite{kogut} and in section IV of this paper.

The universality of the mass spectrum can be understood similarly :
the kinetic term of the nonlinear $\sigma$ model is
naturally derived from the metric on $G/K$ that is induced from
the Killing form on $G$.
We can choose a $\cG$-valued connection $B_\mu$
on the $G$-bundle over
the spacetime manifold
and construct an associated Laplace--Beltrami operator $\nabla^2$.
This procedure leads directly to the  universality of the excitation masses.

\begin{acknowledgments}
We thank Alex Kovner for participating in the early stages of this work,
and Jac Verbaarschot for helpful discussions and comments.
We also thank the U.S. DOE for support through the grant DE-FG02-92ER40716.
\end{acknowledgments}


\begin{thebibliography}{99}
\bibitem{miransky}
V.A. Miransky,
``Dynamical Symmetry Breaking in Quantum Field Theories",
{\sl World Scientific} (Singapore, 1993).

\bibitem{vafa}
C.~Vafa and E.~Witten,
``Restrictions on symmetry breaking in vector-like gauge theories,''
Nucl.\ Phys.\ B {\bf 234}, 173 (1984);
``Eigenvalue inequalities for fermions in gauge theories,''
Commun.\ Math.\ Phys.\  {\bf 95}, 257 (1984);
``Parity conservation in QCD,''
Phys.\ Rev.\ Lett.\  {\bf 53}, 535 (1984).

\bibitem{roman}
R.~Jackiw and S.~Templeton,
``How superrenormalizable interactions cure their infrared divergences,''
Phys.\ Rev.\ D {\bf 23}, 2291 (1981).

\bibitem{poly}
A.~P.~Polychronakos,
``Symmetry breaking patterns in (2+1)-dimensional gauge theories,''
Phys.\ Rev.\ Lett.\  {\bf 60}, 1920 (1988).

\bibitem{pisarski1}
R.~D.~Pisarski,
``Chiral symmetry breaking in three-dimensional electrodynamics,''
Phys.\ Rev.\ D {\bf 29}, 2423 (1984).

\bibitem{stam}
K.~Stam,
``Dynamical mass generation in large N QED in three-dimensions,''
Phys.\ Rev.\ D {\bf 34}, 2517 (1986).

\bibitem{appelquist}
T.~Appelquist, M.~J.~Bowick, D.~Karabali and L.~C.~Wijewardhana,
``Spontaneous breaking of parity in (2+1)-dimensional QED,''
Phys.\ Rev.\ D {\bf 33}, 3774 (1986);
T.~W.~Appelquist, M.~J.~Bowick, D.~Karabali and L.~C.~Wijewardhana,
``Spontaneous chiral symmetry breaking in three-dimensional QED,''
Phys.\ Rev.\ D {\bf 33}, 3704 (1986);
T.~Appelquist, D.~Nash and L.~C.~Wijewardhana,
``Critical behavior in (2+1)-dimensional QED,''
Phys.\ Rev.\ Lett.\  {\bf 60}, 2575 (1988);
D.~Nash,
``Higher order corrections in (2+1)-dimensional QED,''
Phys.\ Rev.\ Lett.\  {\bf 62}, 3024 (1989);
T.~Appelquist, J.~Terning and L.~C.~Wijewardhana,
``(2+1)-dimensional QED and a novel phase transition,''
Phys.\ Rev.\ Lett.\  {\bf 75}, 2081 (1995)
[arXiv:hep-ph/9402320].

\bibitem{cristina}
M.~C.~Diamantini, P.~Sodano and G.~W.~Semenoff,
``Chiral dynamics and fermion mass generation in three-dimensional gauge theory,''
Phys.\ Rev.\ Lett.\  {\bf 70}, 3848 (1993)
[arXiv:hep-ph/9301256].

\bibitem{dagotto}
E.~Dagotto, J.~B.~Kogut and A.~Kocic,
``A computer simulation of chiral symmetry breaking in (2+1)-dimensional QED with N flavors,''
Phys.\ Rev.\ Lett.\  {\bf 62}, 1083 (1989);
``Chiral symmetry breaking in three-dimensional QED with N(F) flavors,''
Nucl.\ Phys.\ B {\bf 334}, 279 (1990).

\bibitem{pennington}
M.~R.~Pennington and D.~Walsh,
``Masses from nothing: A nonperturbative study of QED in three-dimensions,''
Phys.\ Lett.\ B {\bf 253}, 246 (1991);
D.~C.~Curtis, M.~R.~Pennington and D.~Walsh,
``Dynamical mass generation in QED in three-dimensions and the 1/N expansion,''
Phys.\ Lett.\ B {\bf 295}, 313 (1992).

\bibitem{pisarski2}
R.~D.~Pisarski,
``Fermion mass in three-dimensions and the renormalization group,''
Phys.\ Rev.\ D {\bf 44}, 1866 (1991).

\bibitem{gasser}
J.~Gasser and H.~Leutwyler,
``Chiral perturbation theory to one loop,''
Annals Phys.\  {\bf 158}, 142 (1984);
``Chiral perturbation theory: Expansions in the mass of the strange quark,''
Nucl.\ Phys.\ B {\bf 250}, 465 (1985).

\bibitem{weinberg}
S.~Weinberg,
``The Quantum Theory of Fields, Vol. 2: Modern Applications,''
sect.19, {\sl Cambridge Univ. Press} (Cambridge, UK, 1996).

\bibitem{jac1}
J.~J.~M.~Verbaarschot and T.~Wettig,
``Random matrix theory and chiral symmetry in QCD,''
Ann.\ Rev.\ Nucl.\ Part.\ Sci.\  {\bf 50}, 343 (2000)
[arXiv:hep-ph/0003017].

\bibitem{hands}
S.~Hands, J.~B.~Kogut, M.~P.~Lombardo and S.~E.~Morrison,
``Symmetries and spectrum of SU(2) lattice gauge theory at finite  chemical potential,''
Nucl.\ Phys.\ B {\bf 558}, 327 (1999)
[arXiv:hep-lat/9902034].

\bibitem{kst}
J.~B.~Kogut, M.~A.~Stephanov and D.~Toublan,
``On two-color QCD with baryon chemical potential,''
Phys.\ Lett.\ B {\bf 464}, 183 (1999)
[arXiv:hep-ph/9906346].

\bibitem{kogut}
J.~B.~Kogut, M.~A.~Stephanov, D.~Toublan, J.~J.~M.~Verbaarschot and A.~Zhitnitsky,
``QCD-like theories at finite baryon density,''
Nucl.\ Phys.\ B {\bf 582}, 477 (2000)
[arXiv:hep-ph/0001171].

\bibitem{jac2}
D.~Toublan and J.~J.~M.~Verbaarschot,
``Dirac spectra and real QCD at nonzero chemical potential,''
arXiv:hep-th/0208021.

\bibitem{jac3} J.~J.~M.~Verbaarschot, ``What really matters'', talk at the Channel Meeting, Plymouth, August 2002.

\bibitem{dagotto}
E.~Dagotto, F.~Karsch and A.~Moreo,
``The strong coupling limit of SU(2) QCD at finite baryon density,''
Phys.\ Lett.\ B {\bf 169}, 421 (1986).

\bibitem{baillie}
C.~Baillie, K.~C.~Bowler, P.~E.~Gibbs, I.~M.~Barbour and M.~Rafique,
``The chiral condensate in SU(2) QCD at finite density,''
Phys.\ Lett.\ B {\bf 197}, 195 (1987).

\bibitem{kogut2}
J.~B.~Kogut, D.~Toublan and D.~K.~Sinclair,
``The phase diagram of four flavor SU(2) lattice gauge theory at nonzero  chemical potential and temperature,''
Nucl.\ Phys.\ B {\bf 642}, 181 (2002)
[arXiv:hep-lat/0205019];
``SU(2) lattice gauge theory at nonzero chemical potential and  temperature,''
arXiv:hep-lat/0208076.

\bibitem{damgaard}
P.~H.~Damgaard, U.~M.~Heller, A.~Krasnitz and T.~Madsen,
``A quark-antiquark condensate in three-dimensional QCD,''
Phys.\ Lett.\ B {\bf 440}, 129 (1998)
[arXiv:hep-lat/9803012].

\bibitem{CW}
S.~R.~Coleman and E.~Witten,
``Chiral symmetry breakdown In large N chromodynamics,''
Phys.\ Rev.\ Lett.\  {\bf 45}, 100 (1980).

\bibitem{magnea}
U.~Magnea,
``The orthogonal ensemble of random matrices and QCD in three dimensions,''
Phys.\ Rev.\ D {\bf 61}, 056005 (2000)
[arXiv:hep-th/9907096].

\bibitem{peskin}
M.~E.~Peskin,
``The alignment of the vacuum in theories of technicolor,''
Nucl.\ Phys.\ B {\bf 175}, 197 (1980).

\bibitem{witten}
E.~Witten,
``An SU(2) anomaly,''
Phys.\ Lett.\ B {\bf 117}, 324 (1982).

\bibitem{redlich}
A.~N.~Redlich,
``Gauge noninvariance and parity nonconservation of three-dimensional  fermions,''
Phys.\ Rev.\ Lett.\  {\bf 52}, 18 (1984);
``Parity violation and gauge noninvariance of the effective gauge field action in three-dimensions,''
Phys.\ Rev.\ D {\bf 29}, 2366 (1984).

\bibitem{magnea2}
U.~Magnea,
``Three-dimensional QCD in the adjoint representation and random matrix  theory,''
Phys.\ Rev.\ D {\bf 62}, 016005 (2000)
[arXiv:hep-th/9912207].

\bibitem{gilmore}
R. Gilmore,
``Lie Groups, Lie Algebras, and Some of Their Applications,''
{\sl John Wiley \& Sons} (New York, 1974).

\end{thebibliography}
\end{document}